\documentclass[11pt]{article}
\pdfoutput=1
\usepackage{graphicx}
\usepackage{hyperref}
\usepackage{epsfig}
\usepackage{amsmath}
\usepackage{amsthm}
\usepackage{amssymb}
\usepackage{multirow}
\usepackage{color}
\usepackage[nosort]{cite}
\usepackage{hyperref}
\usepackage[fontsize=11.3pt]{scrextend}
\usepackage{braket,mathrsfs,slashed}
\usepackage{float}
\usepackage{setspace}
\usepackage[caption=false]{subfig}
\newlength{\abstractwidth}
\newcommand{\be}{\begin{equation}}
\newcommand{\ee}{\end{equation}}

\renewcommand{\title}[1]{\vbox{\center\bf{\Large{#1}}}\vspace{5mm}}
\renewcommand{\author}[1]{\vbox{\center#1}\vspace{5mm}}
\newcommand{\address}[1]{\vbox{\center\em#1}}

\usepackage{dsfont}

\usepackage[utf8]{inputenc}
\usepackage{rotating}
\usepackage{tikz-feynman}

\usepackage{dsfont}
\usepackage[paper=a4paper,margin=1.5cm]{geometry}
\usepackage[utf8]{inputenc}
\usepackage{rotating}
\usepackage{soul}

\usepackage{mathrsfs} 
\usepackage{bbm}
\usepackage{etoolbox} 
\usepackage{multirow}

\allowdisplaybreaks

\renewcommand\[{\begin{equation}}
\renewcommand\]{\end{equation}}

\newcommand{\ba}{\begin{eqnarray}}
\newcommand{\ea}{\end{eqnarray}}

\hypersetup{pdfstartview=FitV,colorlinks=true,linkcolor=midblue,citecolor=midblue,filecolor=midblue,urlcolor=midblue}

\definecolor{midblue}{rgb}{0,0,0.5}

\begin{document}
	
		\newgeometry{top=3.1cm,bottom=3.1cm,right=2.4cm,left=2.4cm}
		
	\begin{titlepage}
	\begin{center}
		\hfill \\
		\vskip 0.5cm

		\title{Is the information loss problem a paradox?}

			\author{\large Luca Buoninfante$^{a,\,\star}$, Francesco Di Filippo$^{b,\,\dagger}$}
			
			\address{$^a$High Energy Physics Department, Institute for Mathematics, Astrophysics,\\
			and Particle Physics, Radboud University, Nijmegen, The Netherlands\\[1.5mm]
				$^b$Institute of Theoretical Physics, Faculty of Mathematics and Physics, Charles\\ University, V.~Holešovičkách 2, 180 00 Prague 8, Czech Republic\\[1.5mm]
                $^b$Institut fur Theoretische Physik, Goethe Universit\"at,\\ Max-von-Laue-Str.~1, 60438 Frankfurt am Main, Germany}
				\vspace{.3cm}

		\end{center}

\vspace{0.15cm}

\begin{abstract}
The aim of this chapter is twofold. First, we introduce the information loss problem; second, we provide a critical assessment by thoroughly inspecting the assumptions underlying its formulations. In particular, we argue that if we work in the regime of validity of semiclassical gravity and  do not add additional assumptions that are not necessary for the Hawking calculation, the answer to the question in the title is NO. In other words, the black hole evaporation is certainly unitary as predicted by quantum field theory in curved spacetime. However, if additional assumptions are added, such as a universal area bound on the entropy, contradictions may arise even in regimes where we would expect the semiclassical approximation to be valid. We show that a contradiction indeed arises, but not between the laws of semiclassical general relativity and quantum mechanics, but rather between the former and the additional (holographic) requirement of the area limit, according to which an exterior observer describes a black hole as a quantum system whose entropy is bounded by its area. 
\end{abstract}
\vspace{6.5cm}
\noindent\rule{6.5cm}{0.4pt}\\
$\,^\star$
\href{mailto:luca.buoninfante@ru.nl}{luca.buoninfante@ru.nl}\\	
$\,^\dagger$ \href{mailto:francesco.difilippo@matfyz.cuni.cz}{francesco.difilippo@matfyz.cuni.cz}\\[1.5mm]
Invited contribution to appear as Chapter 4 in ``The Black Hole Information Paradox''  (Eds. Ali Akil and Cosimo Bambi, Springer Singapore).

\end{titlepage}

{
	\hypersetup{linkcolor=black}
	\tableofcontents
}

\baselineskip=17.63pt



\newpage

\section{Introduction}

Fifty years ago Hawking wrote his seminal papers on the Hawking radiation~\cite{Hawking:1974rv,Hawking:1975vcx} and for the first time raised the question of whether information is lost due to black hole evaporation, thus giving rise to the notoriously known ``information loss paradox''~\cite{Mathur:2009hf,Buoninfante:2021ijy}. The usual argument is that semiclassical gravity and quantum mechanics are incompatible because the black hole evaporation violates unitarity by evolving an initial pure state into a final mixed state. Despite its simplicity, this statement hides several assumptions and can in fact be proven logically incorrect if we work within the validity regime of the semiclassical approximation and do not introduce additional assumptions that are unnecessary for the standard calculation of quantum field theory in curved spacetime.

The aim of this article is to review different formulations of the information loss ``paradox'' and then critically inspect the various assumptions that are usually made. We will present two different formulations of the problem and especially focus on the modern version which is due to Page~\cite{Page:1993wv,Page:1993up}. We refer to the latter formulation as the entropy problem. This states that a logical contradiction can arise even for macroscopic objects, i.e. for time scales much smaller than the lifetime of the black hole, where we would expect the semiclassical approximation to still be valid. The usual claim is that the unitarity of quantum mechanics clashes with the laws of semiclassical general relativity even in low-energy regimes. However, we disagree with this reasoning and will show that the laws of quantum mechanics and general relativity are certainly compatible, at least as long as the semiclassical approximation does not breakdown. In particular, this is the case if we consider time scales much shorter than the duration of black hole evaporation. What we will carefully argue instead is that a contradiction may arise only when further and generally unjustified assumptions are made. In particular, if the \textit{area limit}~\cite{Almheiri:2020cfm} -- according to which a black hole can be described as a quantum object whose entropy is always bounded by its area -- is imposed as one of the starting assumptions, then a contradiction is found. However, the latter is not between semiclassical gravity and quantum mechanics, but between semiclassical gravity and the additional requirement of the area limit~\cite{Buoninfante:2021ijy}. This represents a radically different formulation of the black hole information loss problem.

The work is organized as follows.
\begin{description}

    \item[\textbf{Sec.~\ref{sec:particle-creation}:}] Before discussing the information loss problem, we find it useful to briefly review some key elements of the physical mechanism underlying the Hawking radiation and clarify some aspects that are sometimes misunderstood or misinterpreted. In particular, we will explain how to define vacuum states at early and late times for an evaporating black hole, why a time-dependent geometry is needed to have particle creation, why the standard heuristic picture of particle-antiparticle pair production is unphysical and incorrect, and  what is the role of the horizon.

    \item[\textbf{Sec.~\ref{sec:standard-formulations}:}] Two different formulations of the information loss problem will be presented: the unitarity problem and the entropy problem. In both scenarios, the underlining assumptions will be critically inspected and it will be clarified that there is no incompatibility between quantum mechanics and semiclassical gravity. In particular, we will explain that, within its regime of validity, semiclassical gravity predicts a unitary evolution of quantum states during the black hole evaporation. 
    The computation from quantum field theory in curved spacetime will imply
    that the number of degrees of freedom contained within a black hole cannot be limited by its area.

    \item[\textbf{Sec.~\ref{sec:personal-view}:}] We will make several important remarks on the information loss problem by emphasizing that contradictions may arise only if the area limit is taken as a starting assumption. We will argue that the combination of established theories, such as quantum mechanics and semiclassical general relativity  in their regimes of validity, does not require and is in fact incompatible with the holographic principle. We will also discuss the open question of if, how and when the information retrieval occurs.

    \item[\textbf{Sec.~\ref{sec:take-home}:}] The key points of our discussion will be concisely summarized and the take-home message of this work will be sharply conveyed with the hope that less misunderstanding will arise in the future about the information loss problem.
    
\end{description}

\textbf{Conventions.} Throughout the article we will work with the mostly plus convention for the metric signature $(-+++)$ and adopt the Natural units ($\hslash=1=c$).

\section{Particle creation}\label{sec:particle-creation}

Before delving into the information loss problem, we believe it is useful to recall the main features of particle creation because its physical mechanism is sometimes misunderstood and misinterpreted. First of all, we should distinguish at least two different types of physical scenarios that could give rise to particle creation: (i) particle creation due to relative observations made by observers in two different frames; (ii) particle creation due to some time-dependent process that takes an initial empty vacuum state into a non-empty final state. An example of the first type is the Unruh effect~\cite{Unruh:1976db} according to which an inertial observer sees an empty vacuum, while an accelerating observer sees a thermal bath. The second type of scenario for particle creation is not frame-dependent but is due to some dynamical process that is able to convert an initial empty vacuum state into a non-empty excited state. This can happen either thanks to some time-dependent potential~\cite{Dunne:1998ni} (e.g. a scattering process or the Schwinger effect) or to some dynamical background geometry (e.g. an expanding universe~\cite{Parker:1968mv,Zeldovich:1971mw} or a collapsing black hole~\cite{Hawking:1974rv,Hawking:1975vcx}). 

Both types of particle creation phenomena are characterized by a flux of energy that is produced in the form of radiation. Energy conservation must still be respected in some way, thus the energy produced must come from somewhere. For example, in the case of the Unruh effect, the energy carried by the radiation that is observed in the non-inertial frame comes from the energy that is needed to keep the observer accelerating, e.g. the work done by the engine to accelerate. Furthermore, the time-dependency of a potential can convert potential energy into emitted radiation. An expanding cosmological background and a collapsing star can give rise to some net redshift effects on the propagating waves whose final states are characterized by larger amplitudes, i.e. by larger occupation numbers. In the latter case, the energy of the emitted radiation comes from a variation over time of the gravitational field perceived by the propagating wave, due to the expansion of the universe or the contraction of a collapsing star.

In the following, we will analyze the main aspects of the mechanism of particle creation due to the collapse of a star into a black hole. After introducing some elements of quantum field theory in curved spacetime and defining the concept of vacuum, we will explain in more detail the physical reason why particles are created in this dynamical geometry, thus it will also become clear why no particle creation can occur in a static (eternal) black hole spacetime. 
Furthermore, we will show that the semiclassical computation of Hawking radiation is consistent with a unitary quantum mechanical evolution once all the outgoing Hawking modes are correctly taken into account.

\subsection{Definition of the vacuum state}\label{sec:Vac_state}

To understand the phenomenon of particle creation, we need to properly define the concept of the vacuum state. For simplicity, we only analyze a free massless scalar field whose dynamical equation is
\begin{equation}
    g^{\mu\nu}\nabla_\mu \nabla_\nu \phi=0\,.
\end{equation}
In flat spacetime, we can decompose the field into positive and negative energy modes, i.e.\footnote{It is worth to mention that the summation over $i$ is a short-hand notation for a more general case of an integral over the angular components of $\textbf{k}_i$ and its modulus $|\textbf{k}_i|=\omega_{\textbf{k}_i}$. Below we will also use the notation $a_{\omega}$ and $f_{\omega}$ to indicate the modes with frequency $\omega.$}
\begin{equation}\label{field-decomposition}
    \phi=\sum_i \left(a_i f_i+a_i^\dagger f_i^*\right)\,,
\end{equation}
where $a_i$ and $a_i^\dagger$ are the so-called annihilation and creation operators, while the mode-functions are given by
\begin{equation}
    f_i\propto e^{-i\omega_{\,\textbf{k}_i}t\pm \textbf{k}_{i}\cdot \textbf{x}}\,,\qquad \left|\textbf{k}_{i}\right|=\omega_{\,\textbf{k}_i}\geq0\,.
\end{equation}
The vacuum state is defined as the state annihilated by the operators $a_i,$ i.e.
\begin{equation}
    a_i\left|0\right>=0\,.
\end{equation}
The problem with this definition, is that it requires a choice for the time and spatial coordinates, that is, in general it is not unique. However, in flat spacetime it can be proven that the vacuum state is Poincaré invariant: every inertial observer would agree on the definition of the vacuum state. This does not mean that there is a unique vacuum choice, even in flat spacetime. In fact, according to the Unruh effect, an accelerated observer measures a thermal bath, while an inertial observer sees the same state as an empty vacuum.

In a generic spacetime, the situation is more involved. In general, there is no notion of preferred observer. Hence, there is no preferred choice of the vacuum state. The problem is alleviated for stationary spacetimes with an everywhere defined killing vector $\xi$. In this case, the killing vector provides a preferred notion of time and thus a way to split the positive and negative frequency modes, i.e. we have eq.~\eqref{field-decomposition} where
\begin{equation}
    \xi^\mu\nabla_\mu f_i=-i\omega_i f_i\,,\qquad\omega_i\geq 0\,.
\end{equation}

The existence of a preferred vacuum choice is also the reason why particle creation cannot happen in a stationary spacetime. In particular, in an eternal black hole spacetime no particle is created.\footnote{It is often claimed that Hawking radiation is present even in an eternal black hole spacetime. However, while energy fluxes may indeed be present, they are not due to particle creation. What happens in this case is that one can define the so-called Hartle-Hawking state according to which both vacua on the past null infinity and the future null infinity are not empty, i.e. there is a pre-existing energy flux flowing from past to future and the black hole is in thermal equilibrium with these fluxes (see e.g.~\cite{Basile:2024oms} for a pedagogical discussion).}

While this is only possible for stationary spacetimes, the physically relevant scenarios correspond to backgrounds that are approximately stationary at early times and, after a non-trivial time-dependent dynamical transient, return to being approximately stationary at sufficiently late times. Therefore, we can  identify \textit{two} timelike vector fields $\xi_{in}$ and $\xi_{out}$ that are killing vectors in the two asymptotic regions and write two different mode decompositions of the field $\phi$ in two different bases~\cite{Birrell:1982ix,Fabbri:2005mw}:
\begin{equation}
     \phi=\sum_i \left(a_i^{in} f_i^{in}+a_i^{in\dagger} f_i^{in*}\right)\,,\qquad     \phi=\sum_i \left(a_i^{out} f_i^{out}+a_i^{out\dagger} f_i^{out*}\right)\,.
\end{equation}
We can then define two vacuum states, one for the $in$ region in the asymptotic past and one for the $out$ region in the asymptotic future:
\begin{equation}
    a_i^{in}\left|in\right>=0\,,\qquad a_i^{out}\left|out\right>=0\,.
\end{equation}
These two vacua only agree if the two killing vectors are equal (e.g.~for stationary spacetimes). Generically, they differ and can be shown to be related by the general relation~\cite{Fabbri:2005mw} 
\begin{equation}\label{eq:rel-in-out}
    \left|in\right>=\left<out|in\right> e^{\frac{1}{2}\sum_{ij}V_{ij}a_i^{out\dagger}a_j^{out\dagger}}
    \left|out\right> \,,
\end{equation}
where $V_{ij}$ are functions of the Bogoliubov coefficients that relate the annihilation and creation operators between two asymptotic regions; however, their explicit expressions are not important for our discussion.

As we will see later, a relation between $\left|in\right>$ and $\left|out\right>$ that is not compatible with~\eqref{eq:rel-in-out} could be a sign of unitarity violation or an indication that we have done something wrong in the way we have treated the outgoing radiation, for example by leaving out some of the outgoing modes. The latter, in fact, is the mistake that is sometimes committed and that leads to an apparent contradiction between the unitarity of quantum mechanics and the validity of semiclassical general relativity.

\subsection{Time dependent geometries}

The origin of the Hawking radiation is to be found in the dynamical nature of the spacetime, given by a contracting star that eventually collapses into a black hole. Let us give an intuitive physical explanation of why particles are created in this situation.  Consider a scalar wave that propagates and enters the star, passes through it and escapes back to infinity. In a static configuration, the wave would reach the future asymptotic region unmodified as any effect during the propagation towards the central object would be exactly compensated by opposite effects when the wave propagates away from it.  However, if the star is collapsing, this is no longer true. It is possible to show that the wave would emerge with a net redshift and a net amplification~\cite{Brout:1995rd,Fabbri:2005mw}. This physically means that the state of the outgoing wave is characterized by a larger occupation number as compared to the initial incident one. Therefore, an asymptotic observer on $\mathcal{I}^+$ would detect a flux of energy, i.e. Hawking radiation. We now provide some more details on the spectrum of the radiation.

\begin{figure}[t]
    \centering
    \includegraphics[width=0.3\linewidth]{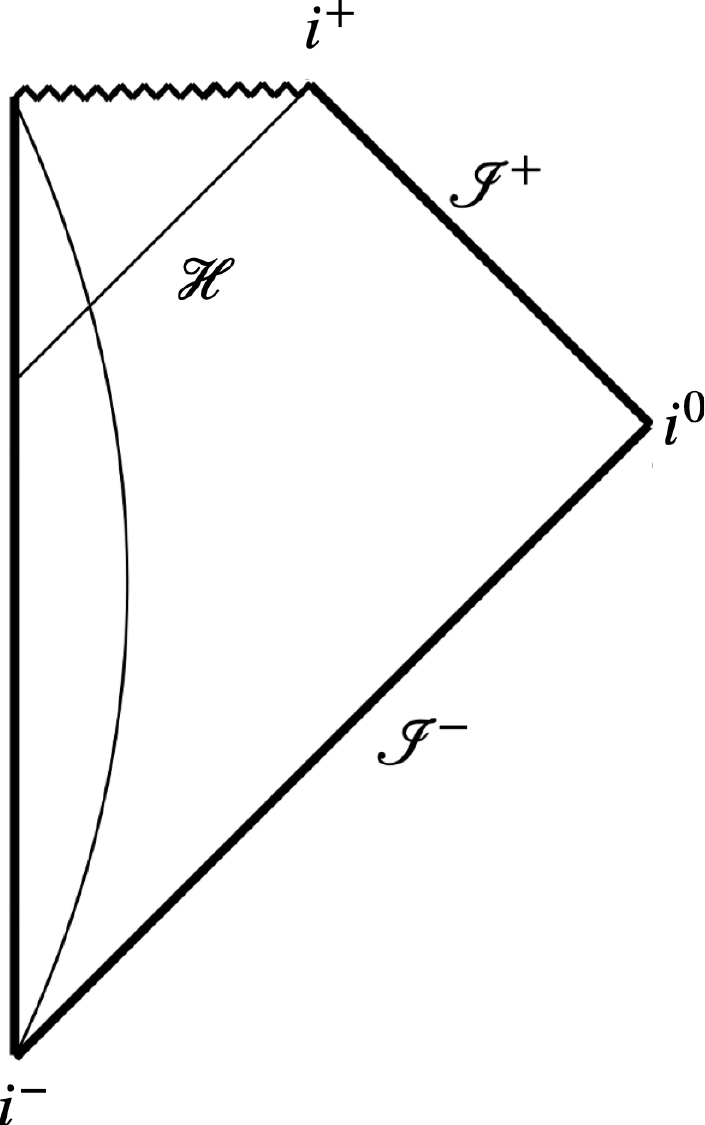}
    \caption{Penrose-Carter diagram of a black hole formed via gravitational collapse. }
    \label{fig:Grav_collapse}
\end{figure}

The Penrose-Carter diagram of a spherically symmetric black hole formed by gravitational collapse is depicted in Fig.~\ref{fig:Grav_collapse}. If we start with a vacuum state on $\mathcal{I}^-$ which is empty, the same state on $\mathcal{I}^+$ will have a non-trivial matter content. The seminal Hawking computation tells us that the expectation value of the particle number operator on $\mathcal{I}^+$ is~\cite{Hawking:1974rv,Hawking:1975vcx}
\begin{equation}
   \left<in\right| N^{out}_\omega \left|in\right>=\frac{\Gamma_{l,\omega}}{e^{8\pi GM\omega}-1}\,,
   \label{eq-spectrum}
\end{equation}
where we have introduced the number operator $N^{out}_\omega=a^{out\dagger}_{\omega}a^{out}_\omega.$
The expression in eq.~\eqref{eq-spectrum} is the same distribution of the thermal spectrum of radiation for bosons. This implies that an asymptotic observer on $\mathcal{I}^+$ sees a flux of radiation that follows a Planckian distribution with temperature inversely proportional to the mass of the black hole, i.e.  
\begin{equation}
T_{H}=\frac{1}{8\pi G M}\,.
\end{equation}
Furthermore, the quantity $\Gamma_{l,\omega}$ is the so-called gray-body factor that physically represents the transmission coefficient of the wave through the potential that is peaked around the radius of the photosphere; $l\in \{0,1,2,\dots\}$ is the angular momentum number associated with the spherical harmonic decomposition of the scalar field. This means that a fraction $1-\Gamma_{l,\omega}$ is backscattered by the potential, falls back into the black hole and does not reach the asymptotic region at infinity.

To confirm that the radiation is truly thermal, one has to check that different quanta with frequencies $\omega$ and $\omega'$ on $\mathcal{I}^+$ are uncorrelated. A direct computation shows that this is the case~\cite{Wald:1975kc} (see also~\cite{Fabbri:2005mw} for a pedagogical analysis). In particular, we have
\begin{equation}
\left<in\right|N^{out}_{\omega}N^{out}_{\omega'}\left|in\right>= \left<in\right|N^{out}_{\omega}\left|in\right> \left<in\right|N^{out}_{\omega'}\left|in\right>\qquad \text{for}\quad \omega\neq \omega'\,,
\end{equation}
which indeed means that there is no correlation between the modes on $\mathcal{I}^+$. Therefore, starting from a pure state on $\mathcal{I}^-$ we obtain a mixed state on $\mathcal{I}^+$, in contradiction with~\eqref{eq:rel-in-out} which would instead tell us that the in-state is pure also for observers in the out region as it can be written as a linear combination of pure states. Given that unitary operators must evolve pure states into pure states, we might be temped to conclude that we have reached a paradox.
However, this conclusion would be too hasty. As we will explain below, this apparent contradiction is caused by the fact that we have left out some of the outgoing modes, in particular those that do not reach $\mathcal{I}^+$ but that instead cross the event horizon and enter the black hole~\cite{Wald:1975kc}.

It is worth mentioning that Hawking's original computation neglects the backreaction, i.e. the effect of the emitted radiation on the spacetime geometry. Approximating the black hole as a infinite reservoir\footnote{The black hole can be approximated as an infinite reservoir by taking the double limit $M\rightarrow \infty$, $G\rightarrow 0,$ with $GM$ kept fixed. In this way, the black hole mass is infinite and the right-hand side of the Einstein's equation vanishes (i.e.  backreaction is neglected) but, at the same time, the black hole radius $2GM$ and its evaporation rate $\sim 1/(GM)^2$ are finite.} is definitely a good approximation for a certain interval of time, but eventually the radiated energy would become comparable to the mass of the black hole so that the assumption of a fixed background geometry (i.e. no backreaction) would not hold anymore~\cite{Barcelo:2010pj,Dvali:2015aja,Agullo:2024nxg}.

\textbf{Remark.} To clarify the language we use in this chapter, let us emphasize that with the expression ``semiclassical approximation'' we refer to a low-energy regime in which general relativity can be considered a reliable effective field theory, that is, the characteristic energy scales are lower than the effective-field-theory cutoff which we take to be equal to the Planck mass, in particular the black hole mass has to be larger than the cutoff. This means that matter falls into a black hole according to the laws of general relativity, for example at the horizon scale nothing special happens due to the equivalence principle. Furthermore, time scales of the order of half the black-hole lifetime can still be described within the semiclassical approach, the backreaction of Hawking radiation on the black hole geometry does not have to be neglected and the radiation does not have to be completely thermal. Indeed, backreaction and non-thermal corrections can be consistently taken into account in the adiabatic approximation $\dot{T}_{H}/T_H^2\ll 1\Leftrightarrow 1/GM^2\ll 1$ and within the same framework of quantum field theory in curved spacetime~\cite{Barcelo:2010pj}. In our definition of semiclassical approximation, quantum-gravitational effects can be present as long as the black-hole mass is  bigger than the Planck mass, for example we can have gravitons interacting in an evaporating black-hole spacetime.

Before concluding this section, we want to clarify some additional aspects regarding the physical mechanism of Hawking radiation.

\subsection{Heuristic picture and its limitations}

The heuristic argument that is often invoked for understanding Hawking evaporation is quite simple and is worth recalling it here to explain why it is physically incorrect. 

It is sometimes said that in quantum mechanics the vacuum is not empty, but it is populated by virtual particle-antiparticle fluctuations. When a pair is created near a black hole horizon, it is possible that the two particles are separated by tidal forces and become real. The positive energy particle escapes to infinity, while the negative energy one enters the horizon and is destroyed at the singularity. For an asymptotic observer, this process is equivalent to a flux of positive energy leaving the black hole. This heuristic explanation of the origin of Hawking radiation is widely spread due to its simplicity. However, while it has some merit in partly capturing the physics at play, it falls short in several regards and can lead to confusions.

One of the main limitations of this heuristic argument is that it makes no mention to the time-dependence of the spacetime and it seems to be applicable to static geometries as well. However, we have seen in Sec.~\ref{sec:Vac_state} that particle creation only happens when the spacetime is time-dependent as only in that case the two vacuum states defined in the asymptotic regions do not agree.
In fact, even if the early- and late-time geometries are stationary, the root of Hawking radiation lies in the time-dependence of the geometry that is necessary to dynamically form the black hole. Furthermore, while the heuristic picture requires that the ingoing and outgoing particles are equally correlated throughout the entire evaporation process, the rigorous semiclassical gravity result shows that the outgoing modes are mostly correlated only with the early-time ingoing~modes.

From the discussion above, it should also be clear that the cause of Hawking radiation is not to be found in the presence of a horizon. In fact, it is possible to have particle production even without the formation of a horizon. However, if the horizon does not form, the geometry eventually relaxes into a static configuration and the radiation stops. The formation of the horizaon makes the evaporation continue forever (or at least for a very long but finite time if we take into account the backreaction according to which the mass has to decrease). Finally, the presence of a horizon leads to universal radiation, whose the thermal distribution does not depend on the details of the gravitational collapse~\cite{Hawking:1975vcx} (see also ~\cite{Basile:2024oms,Fabbri:2005mw} for pedagogical explanations).

In the remaining of this chapter, we will clarify that information loss problems may arise but only if some extra ingredients and assumptions are added beyond the framework of semiclassical general relativity.

\section{Standard formulations of the information loss problem}\label{sec:standard-formulations}

In the previous section, we have seen that a pure state on $\mathcal{I}^-$ does not evolve into a pure state on $\mathcal{I}^+$. Given that a unitary evolution must evolve pure states into pure states, we might be temped to conclude that this result, derived within semiclassical gravity, is paradoxical. However, this is not the case. 
In fact, contrary to $\mathcal{I}^-$, the null region $\mathcal{I}^+$ is not a Cauchy hypersurface\footnote{Note that this is an abuse of nomenclature. A Cauchy hypersurface must be acausal, while $\mathcal{I}^-$ and $\mathcal{I}^+$ are null and, hence, neither is a Cauchy hypersurface. What we mean here, is that $\mathcal{I}^-$ intercepts all the modes, while $\mathcal{I}^+$ does not.}. Therefore, we do not need to recover a pure state on $\mathcal{I}^+$ simply because some modes can be lost inside the event horizon $\mathcal{H}$. A unitary evolution operator must evolve a pure state on $\mathcal{I}^-$ into a pure state on $\mathcal{I}^+\cup\mathcal{H}$. This is indeed the case. The framework of semiclassical gravity, in particular eq.~\eqref{eq:rel-in-out}, tells us that the $in$ state is a pure state even when expressed in terms of particle states in the $out$ region, if all the outgoing modes $a^{out\dagger}_i$ are correctly taken into account~\cite{Wald:1975kc}.
Hence, unitarity is not violated and there is no logical contradiction. 

What still needs to be understood is where the correlations are. We have discussed that the radiation on $\mathcal{I}^+$ is uncorrelated. Therefore, the only possibility is that modes on $\mathcal{I}^+$ are correlated with modes that enter the horizon. There are subtleties to verify this possibility as the geometry in some regions of the event horizon is time-dependent and thus it is not easy to provide a well-defined notion of particles. However, there are reasonable ways to define the modes that enter the horizon and, indeed, explicit computations show that the late-time modes on $\mathcal{I}^+$ are maximally entangled with the modes that enter the horizon at early times~\cite{Wald:1975kc}. If these correlations are properly taken into account, then the relation in eq.~\eqref{eq:rel-in-out} can be shown to be true, which also means that an initial pure state $\left|in\right>$ evolves into a final pure state $\left|out\right>$.

So far so good, but where is the paradox if there is a way to understand the correlations between ingoing and outgoing quanta and still be consistent with unitarity? The short answer is that there is no paradox. In fact, logical contradictions can arise only if additional assumptions are introduced.  The standard formulations of the information loss problem require some extra ingredients that go beyond what is necessary for the semiclassical computation of the Hawking radiation. In the reminder of this section we will carefully explain what are the ingredients that are added to the semiclassical description and that could lead to contradictions; we will closely follow the discussion in Ref.~\cite{Buoninfante:2021ijy}.

\subsection{Complete evaporation and unitarity problem}

The original information loss problem is usually stated as a contradiction between the unitarity of the evolution operator in quantum mechanics and the validity of the semiclassical approximation until the end of black hole evaporation. However, this incompatibility is true only if the black hole fully evaporates leaving behind an empty regular spacetime plus radiation. More precisely, what is usually referred to as ``unitarity problem'' can be formulated as the impossibility of coexistence of the following three assumptions:
\begin{enumerate}
    
    \item[A1:] The evolution operator in quantum mechanics is unitarity, in particular pure states evolve into pure states;

    \item[A2:] Semiclassical general relativity is valid during the entire process of black hole evaporation, in particular the emitted radiation on $\mathcal{I}^+$ is in a thermal state;

    \item[A3:] The black hole evaporates completely, leaving behind a regular empty spacetime, plus radiation.
    
\end{enumerate}

\begin{figure}
    \centering
    \includegraphics[width=0.3\linewidth]{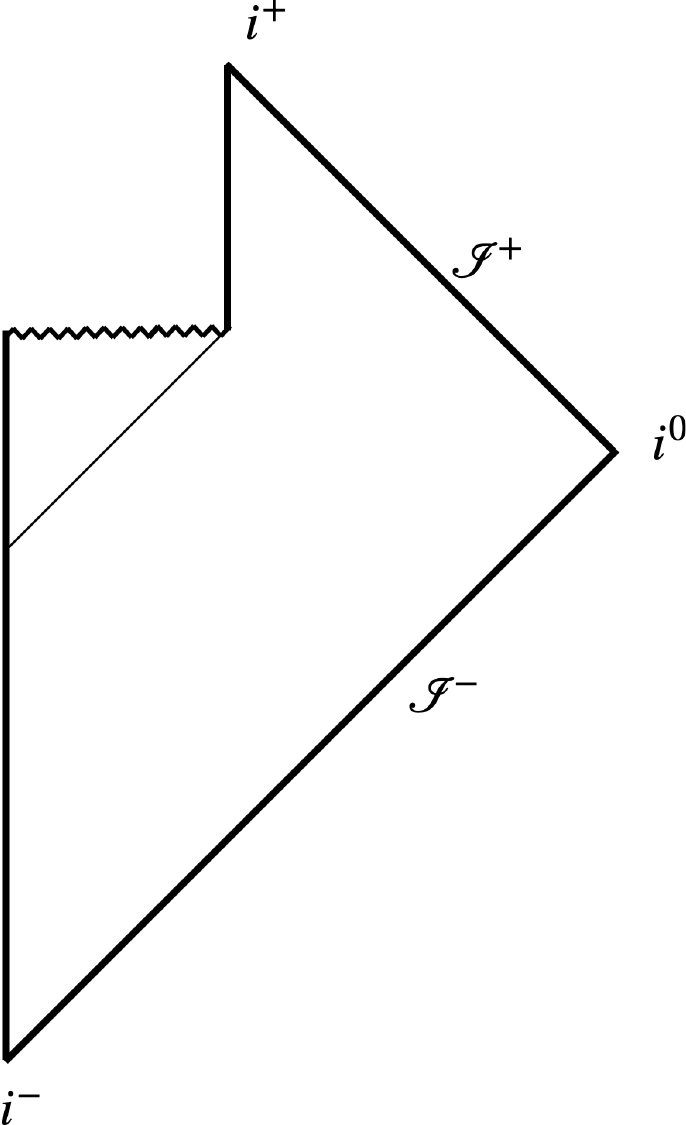}\quad\,\, 
    \includegraphics[width=0.3\linewidth]{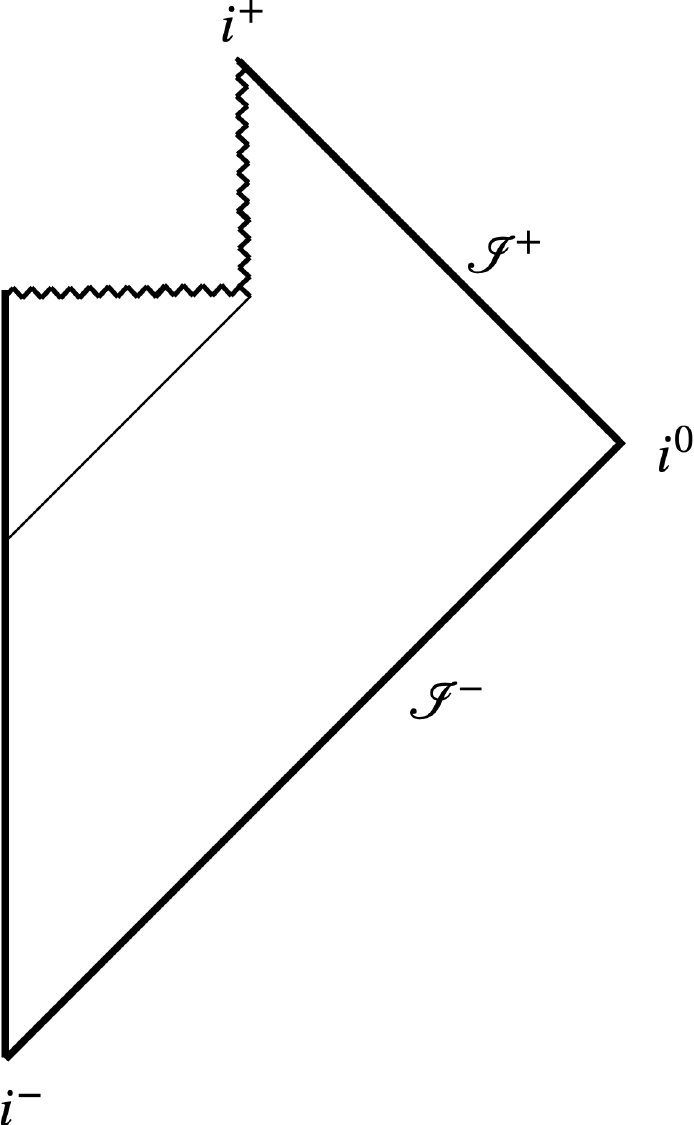}\quad\,\,
    \includegraphics[width=0.3\linewidth]{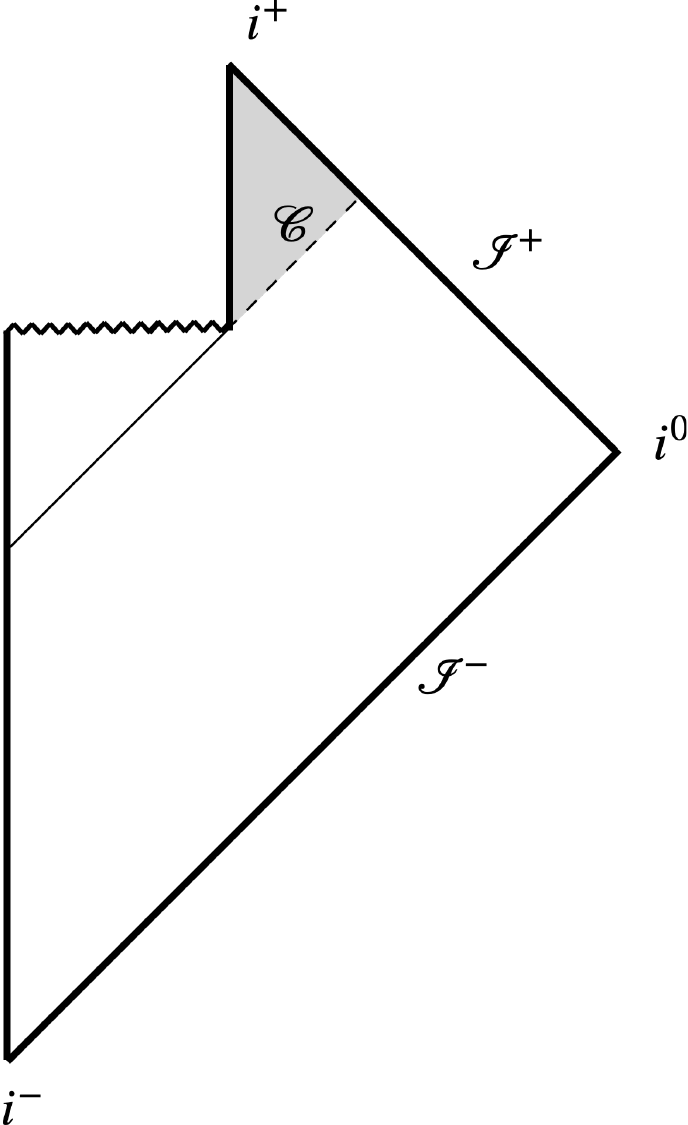}
    \caption{Penrose--Carter diagrams of different possibilities for the endpoint of the evaporation. The left diagram corresponds to the case in which the black hole completely evaporates and leaves an empty regular spacetime plus radiation. The middle diagram represents a scenario in which the endpoint of black hole evaporation corresponds to a naked singularity plus radiation. The right diagram emphasizes the fact that the shaded gray region is in the causal future of the singularity, thus the dashed 
    line $\mathcal{C}$ represents a Cauchy horizon.}
    \label{fig:Grav_collapse-evap}
\end{figure}

If the black hole fully evaporates according to semiclassical gravity and leaves a regular empty spacetime (see the first diagram in Fig.~\ref{fig:Grav_collapse-evap}), then the final state of the process is mixed: the radiation is in a mixed state and nothing is left to purify it. This then implies that unitarity is violated because an initial pure state would evolve into a mixed state.\footnote{It is not important that the initial state is pure. Indeed, even if the initial state is mixed, the unitarity problem would still arise because such a mixed state would be evolved into another mixed state that is not the one predicted by a unitary evolution.} However, we should emphasize that if any of the three assumptions is given up, then no problem or paradox arises. For example, if A1 is not required, then the evolution operator is allowed to be non-unitary. If A2 is dropped, then we cannot trust semiclassical gravity until the endpoint of the evaporation and the final state of the Hawking radiation does not have to be mixed. If A3 is not assumed, it may happen that the black hole evaporation leaves a naked singularity (like in the middle diagram in Fig.~\ref{fig:Grav_collapse-evap}) or even some remnant that could be entangled with the outgoing radiation and eventually purify the final joint state. Therefore, this original formulation of the information loss problem only arises if A1, A2 and A3 are simultaneously assumed. It is generally believed that A2 cannot be true because quantum-gravity effects are expected to become relevant close to the endpoint of black hole evaporation.

\subsection{Entropy problem}

The previous formulation of the information loss problem is mainly related to the physics of the endpoint of the evaporation. However, there exists a stronger formulation according to which a paradox could arise at time scales much shorter than the black hole lifetime. This is originally due to Page~\cite{Page:1993wv,Page:1993up} and is sometimes called ``entropy problem''. It is formulated as an incompatibility among the following three assumptions:
\begin{enumerate}
    
    \item[B1:] The evolution operator in quantum mechanics is unitarity, in particular pure states evolve into pure states;

    \item[B2:] Semiclassical general relativity is a valid low-energy effective field theory as long as the black mass is much larger than the Planck mass;

    \item[B3:] As seen from an outside observer, a black hole can be described as a quantum object whose entropy is bounded by $A/4G,$ where $A$ is the horizon area.\footnote{For the purposes of the discussion it is irrelevant what type of horizon we are referring to when talking about the area limit as in the standard picture the area of the event horizon is very close to the area of the trapping horizon.}
    
\end{enumerate}

As compared to the previous formulation, we have ${\rm A1}={\rm B1}$, but the second assumption is now weaker, i.e. ${\rm B2}\subset {\rm A2}$. This means that the semiclassical approach can be considered reliable only for black holes whose mass is much larger than the Planck mass, and which are therefore still in an evaporation stage far from the endpoint. The new important ingredient is the assumption B3 which is telling us that the total thermodynamic entropy of the black hole\footnote{Here, by the expression ``total thermodynamic entropy'' we mean the thermodynamic entropy of the black hole taking into account the degrees of freedom of both the geometry and the matter inside the black hole, i.e. the horizon and the bulk degrees of freedom, respectively.} $S_{\rm th}$ is bounded by the area, i.e. $S_{\rm th}\leq A/4G$. We now explain why B1, B2, B3 cannot be simultaneously compatible.

\begin{figure}[t!]
    \centering
    \includegraphics[width=0.55\linewidth]{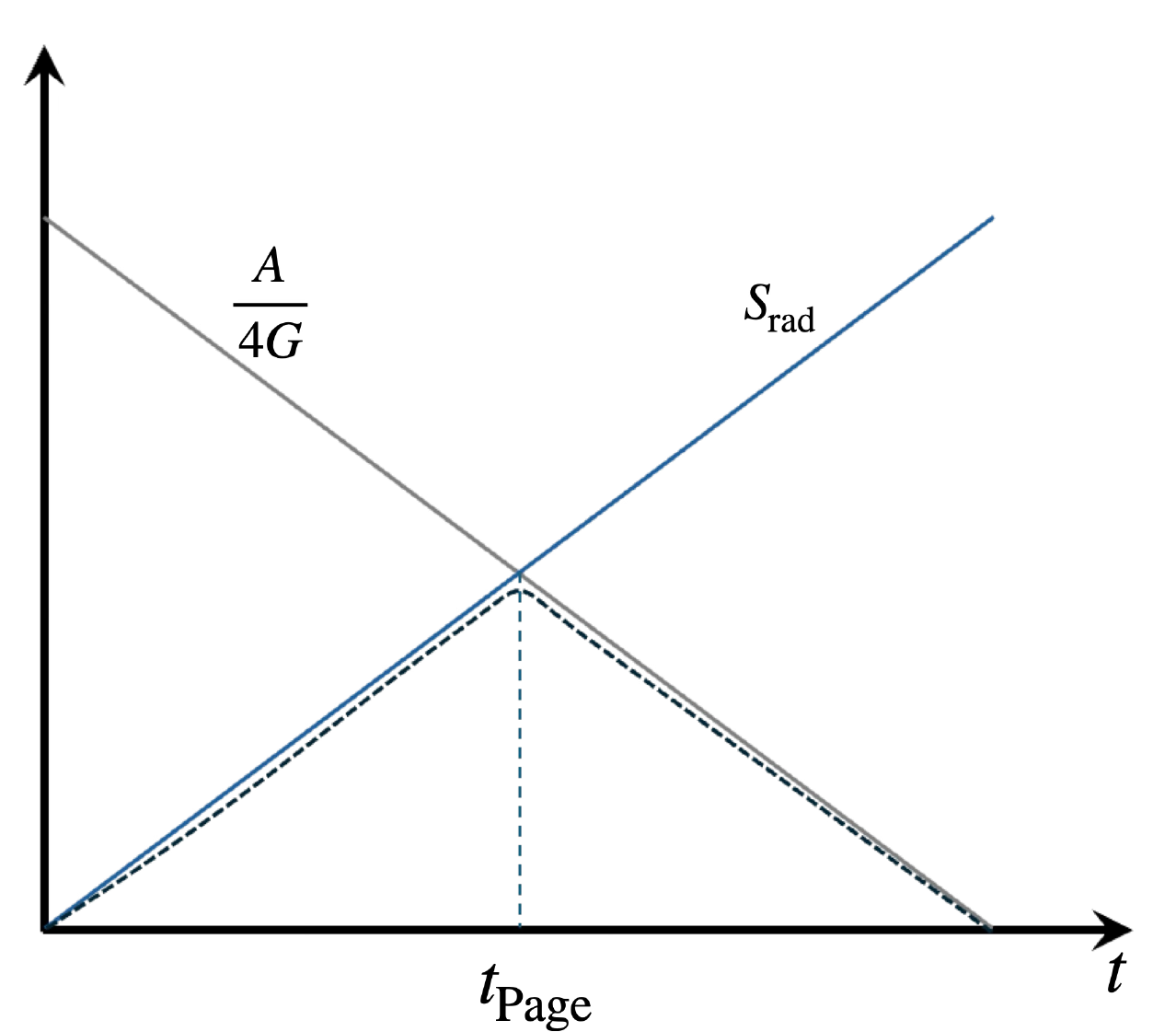}
    \caption{Behavior of various entropies. The gray line is the area $(A/4G)$ of the black hole that decreases in time; the blue line corresponds to the entanglement entropy of the outgoing radiation ($S_{\rm rad}$) which increases according to the prediction of semiclassical gravity; the dashed black line is the entropy of the outgoing radiation that follows the Page curve if the assumption B3 on the area limit is required as an additional hypothesis.}
    \label{fig:Page_Curve}
\end{figure}

Let us call $\mathbb{H}_{\rm bh}$ and $\mathbb{H}_{\rm rad}$ the Hilbert spaces of the black hole and the emitted radiation, respectively. If the initial state before the formation of the black hole was pure, then B1 implies that the joint state of black hole plus radiation must also be pure, that is, the state $\left|\psi \right\rangle \in \mathbb{H}_{\rm bh}\otimes \mathbb{H}_{\rm rad}$ must have zero von Neumann entropy. Tracing over the degrees of freedom of the black hole, we obtain the density matrix for the subsystem of radiation whose von Neumann entropy $S_{\rm rad}$ coincides with that of the black hole subsystem which we call $S_{\rm bh}$. During the evaporation the area decreases. On the other hand, the entropy $S_{\rm rad}$ of the emitted radiation (and similarly $S_{\rm bh}$) will continue to increase according to the semiclassical computation which can be trusted as long as B2 is true. Therefore, there exists a time scale\footnote{It is worth to mention that different notions of time are possible. Here we are considering a time function whose vector field is orthogonal to the Cauchy foliation.} -- denoted by $t_{\rm Page}$ and known as Page time -- after which the von Neumann entropy of the radiation (and thus of the black hole subsystem) exceeds $A/4G$; see Fig.~\ref{fig:Page_Curve}. Consequently, B3 implies that $S_{\rm rad}$ becomes larger than the total thermodynamic entropy of the black hole. In formula we have 
\begin{equation}\label{ineq-Srad>A}
t>t_{\rm Page}\quad \Rightarrow\quad S_{\rm rad}=S_{\rm bh}>\frac{A}{4G}\geq S_{\rm th}\,.
\end{equation}

From the the last inequality, it follows that for time scales $t>t_{\rm Page}$ the joint state $\left|\psi \right\rangle \in \mathbb{H}_{\rm bh}\otimes \mathbb{H}_{\rm rad}$ becomes mixed, thus contradicting the hypothesis B1 and giving rise to a paradox.

As a way to solve this puzzle and preserve unitarity, it has been argued that the von Neumann entropy of Hawking radiation must start decreasing at time scales of the order of $t\sim t_{\rm Page}$ and tend to zero at the end of evaporation~\cite{Page:1993wv,Page:1993up,Penington:2019kki,Almheiri:2019hni,Almheiri:2020cfm}. In this way the von Neumann entropy of radiation would be purified consistently with the requirement of unitarity in B1 and would follow the so-called Page curve shown in Fig.~\ref{fig:Page_Curve}. However, this type of resolution clearly requires giving up B2 because new physics beyond the standard semiclassical gravity result is needed at time scales much shorter than the black hole lifetime in order to obtain the Page curve.

It is worth mentioning that some attempts have been made to obtain a behavior compatible with the Page curve by simply taking into account the backreaction of the emitted radiation. However, it seems that this may not be sufficient~\cite{Mathur:2009hf}.


\subsection{Violation of semiclassical gravity} 

We have inspected the standard formulations of the information loss problem and noticed that a logical contradiction only arises once extra assumptions are added. In particular, the first formulation of the problem needs additional requirements on the endpoint of the evaporation. Therefore, it is not particularly worrisome since it is reasonable to believe that we cannot trust the semiclassical approximation in that regime where quantum-gravity effects should not be negligible. 

The second formulation of the problem is usually considered more problematic as the validity of semiclassical gravity is only assumed in regimes that are still far from the endpoint of black hole evaporation. However, in this case a limit on the numbers of degrees of freedom that can be contained inside a  black hole is conjectured. This corresponds to an additional requirement that is not needed for the standard semiclassical computation of Hawking radiation and actually cannot even be obtained by a self-consistent computation within the model. Interestingly, this second formulation violates the spirit of semiclassical gravity from the very beginning.
In fact, the consistent way to deal with the evolution of a quantum state in curved spacetime would be to choose a Cauchy foliation and consider the evolution of the state from one spacelike hypersurface to the other. In doing so, we are forced to consider Cauchy hypersurfaces that enter the event horizon. By correctly taking into account the outgoing modes on the region of the hypersurface inside the black hole we can make sure that unitarity is satisfied. Indeed, semiclassical gravity predicts that a pure state evolves into a pure state, but it does so in a way that violates the entropy limit as shown in eq.~\eqref{ineq-Srad>A}.

There is no local observation that can detect the presence of an event horizon, and the equivalence principle implies that nothing special should happen there. However, the requirement B3 implicitly assumes that the number of degrees of freedom on the regions of the Cauchy hypersurfaces that are  inside the black hole is  limited by the area of the horizon. This means that new physics beyond semiclassical general relativity is needed to explain the entropy limit on $S_{\rm bh}$ and $S_{\rm rad}$~\cite{Buoninfante:2021ijy,Rovelli:2017mzl,Rovelli:2019tbl}.

\section{Discussion}\label{sec:personal-view}

The information loss problem is often formulated as the incompatibility between the assumption of unitary evolution in quantum mechanics, in particular the fact the pure states evolves into pure states, and the validity of semiclassical gravity which predicts that the entanglement entropy of the Hawking radiation can become larger than the area limit. However, after a careful analysis of the physical assumptions and logical implications, we have shown that in fact unitarity and semiclassical gravity are compatible, i.e. B1 and B2 can coexist. A contradiction arises only when the area limit on the total black hole entropy (B3) is required.

We will now make some important observations that also reflect our view on the information loss problem and on the fact that the area limit cannot be satisfied in the standard framework of semiclassical gravity, but requires the introduction of new physics. 

\medskip

\textbf{Area limit for eternal geometries.} The first point worth discussing is that there are several arguments in favor of the area limit on the number of degrees of freedom contained in a black hole~\cite{Almheiri:2020cfm}. However, to our knowledge, all these arguments are made for stationary eternal geometries. The causal structure is very different from that of a dynamical black hole. It would therefore be essential to quantify the evidence for the area limit in dynamical spacetimes before claiming that B3 can be a justified assumption.

\medskip

\textbf{Gravitational entropy in dynamical configurations.} In relation to the previous point, a major difficulty lies in the fact that it is still not known how to give a definition of gravitational entropy in dynamical configurations.  Let us consider a star that collapses and forms a black hole. It is widely accepted that the star will have an entropy which scales with its volume. On the other hand, it is equally widely accepted that once the black hole is formed the entropy will scale as its area. Perhaps, understanding the transition from a volume law to an area law could shed light on understanding B3 and thus whether there is an information loss problem. However, we believe that the existence of an area limit on the total black hole entropy will still need some new physics beyond semiclassical gravity.

\medskip

\textbf{Decoupling between horizon and bulk degrees of freedom.} An intriguing possibility for the transition from a volume law to an area law could simply be related to the fact that the bulk degrees of freedom decouple and become inaccessible from the outside. Therefore, external observers can describe thermodynamical properties of the black hole simply by looking at the horizon degrees of freedom whose number is limited by $A/4G$. In this picture, the degrees of freedom of the bulk decouple but do not disappear, in particular they still contribute to the von Neumann entropy of the black hole. However, even in this case the total entropy of the black hole would still not be limited by the area and the hypothesis B3 would still not be justified.

\medskip

\textbf{Role of the singularity.} Within the context of semiclassical gravity, a spacetime singularity forms. The theory cannot predict the fate of anything that reaches the singularity. In particular, information falling into the singularity will be destroyed and leave the manifold from the point of view of classical or semiclassical observers. This suggests that to fully understand the problem of information loss in evaporating black holes we must be able to provide a description of the singularity and predict the fate of observers who fall into it. On the other hand, solving the information loss problem by keeping the area limit and assuming a Page curve-like behavior for the entanglement entropy of Hawking radiation would imply that the information can be retrieved without understanding the fate of the singularity.

\medskip

\begin{figure}[t!]
    \centering
    \includegraphics[width=0.4\linewidth]{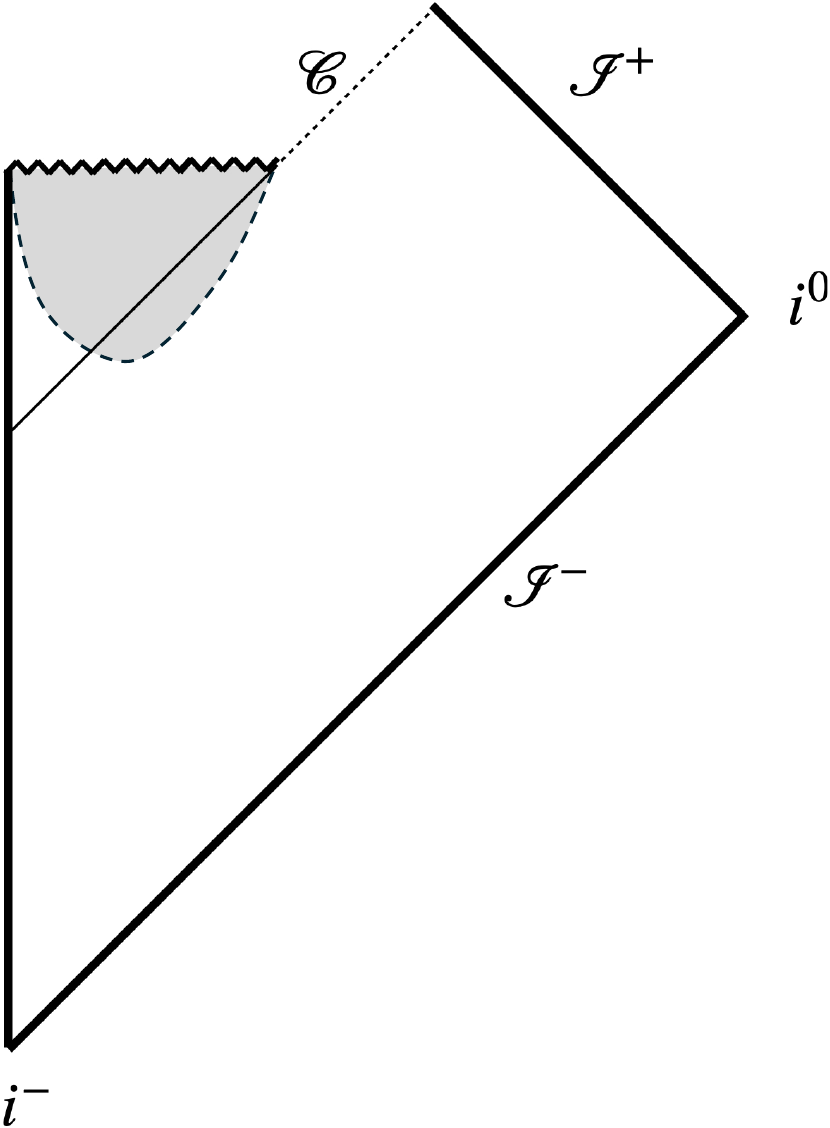}
    \hspace{0.15\linewidth}
    \includegraphics[width=0.3\linewidth]{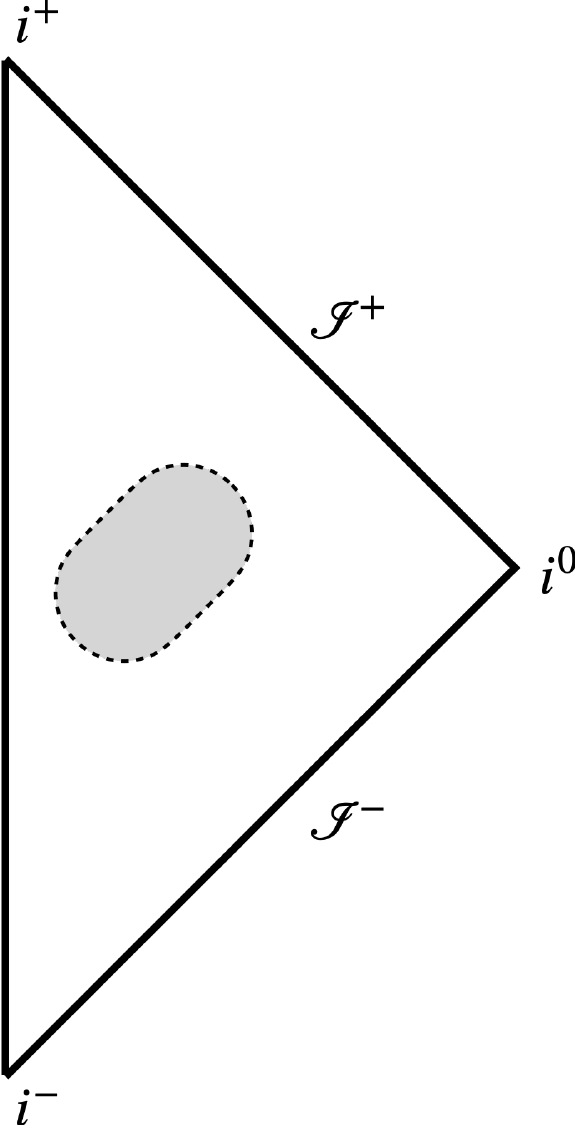}
    \caption{Causal structures of two evaporating black holes. On the left, a singular black hole evaporates and leaves a spacetime with a Cauchy horizon. We do not draw the spacetime beyond the Cauchy horizon because points in that region would lie in the causal future of the singularity. The spacetime has both an event horizon (black solid null line) and a trapping horizon (dashed line enclosing the shaded region). On the right, a regular black hole with no singularity evaporates completely and leaves a regular spacetime. There is no event horizon. The trapping horizon (dashed line enclosing the shaded region) closes on itself. In the left spacetime, we expect information to be lost into the singularity or hidden behind the Cauchy horizon. In the right spacetime, we expect the information to be completely recovered at the end of evaporation.}
    \label{fig:evet_vs_trap}
\end{figure}

\textbf{Event or trapping horizon.} Related to the previous point, we should distinguish between event horizons and long-lived dynamical horizons.  If we form a true event horizon, anything that crosses it will be lost from the point of view of classical asymptotic observers~\cite{Visser:2014ypa,Unruh:2017uaw}. On the other hand, if the horizon is a long-lived trapping horizon, the information should eventually leak out; see Fig.~\ref{fig:evet_vs_trap} for an illustration of these two possibilities. However, event horizons are teleological in nature~\cite{Visser:2014ypa} and a classical observer cannot detect their presence in finite time. With reference to Fig.~\ref{fig:evet_vs_trap}, the situation is slightly different as observer on $\mathcal{I}^+$ can detect the presence of an event horizon if they reach the Cauchy horizon. However, this is only possible at the end of the evaporation which is way after the Page time.
On the other hand, retrieving the information in a way that is consistent with the area limit (i.e. with the Page curve) would require knowing that an event horizon is absent, but the semiclassical gravity approach is not able to provide this knowledge in a regime still very far from the endpoint of evaporation. Note that there are studies on particle creation in the presence of dynamical horizons (while an event horizon is not required)~\cite{Agullo:2024nxg}: in this case one can show that the information cannot leak out of the trapped region within the regime of validity of semiclassical gravity.
Therefore, an information retrieval that is consistent with the area limit seems to require new physics beyond the semiclassical approach already in the low-energy regimes far from the endpoint of evaporation.

\medskip

\textbf{Holography vs semiclassical gravity.} The existence of an area limit on the entropy of a system is usually advocated by the holographic principle~\cite{Bousso:2002ju}. A claim that is often made is that holography should be a fundamental property of a consistent theory of quantum gravity, and thus several researchers consider this to be strong support for the area limit. However, conjecturing holography is the same as requiring the additional assumption B3 in the second formulation of the information loss problem discussed in the previous section. We therefore know that holography will require new physics beyond semiclassical gravity already in the low-energy regime, that is, on time scales much shorter than the lifetime of the black hole.

\section{Take-home message}\label{sec:take-home}

We are now in the position to answer the question in the title of this chapter ``Is the information loss problem a paradox?". 
A paradox is a logical impossibility such as a self-contradictory statement. In this regard, the information loss problem is \textit{not} a paradox. The popular formulation of the problem often expresses it as an incompatibility between the low-energy description of semiclassical gravity and the unitary evolution of quantum states. Given that the equations of motion of semiclassical gravity are built to lead to a unitary evolution, such an incompatibility would imply a paradoxical conclusion. However, we have seen that the formulation of the problem requires extra assumptions that cannot be derived within the semiclassical model. Therefore, the answer to the question is NO, there is no paradox. 

The information loss problem is not an incompatibility between quantum mechanics and semiclassical general relativity, but it is an incompatibility between the validity of semiclassical general relativity and the assumption of the area limit on the total entropy of a black hole. This is not a paradox, but the community should investigate why these two well-motivated hypotheses are incompatible. Once all the outgoing modes of the radiation are properly taken into account both on $\mathcal{I}^+$ and on the event horizon, then one can show that indeed an initial pure state is evolved into a final pure state. The mistake that is sometimes made is to neglect the outgoing modes that cross the event horizon; doing this would obviously give rise to an inconsistent quantum evolution which violates unitarity. Hence, the growth of the entanglement entropy of the Hawking radiation is consistent with unitarity but does not respect the area limit.

Although there is no paradox, there are still open questions about black hole evaporation. These are whether, how, and when information retrieval occurs. We stressed that the presence or absence of the singularity is crucial for this question. If the singularity is there, and thus an event horizon is present, then all the information which falls into the black hole will be crashed into the singularity or hidden behind the Cauchy horizon, that is, in both cases it will be lost for an outside observer. This is not a paradox or a violation of unitarity, but only a consequence of the fact that part of the information has leaked out of the region of spacetime observable by an external observer. If instead the singularity is not there and the event horizon is replaced by a trapping horizon, then all the information can be retrieved. This means that quantum-gravity effects that aim to solve the singularity problem might also shed light on the question of information retrieval in black hole evaporation. 

On the other hand, if one postulates an area limit on the total entropy of the black hole, then information retrieval occurs automatically thanks to the Page curve-like behavior of the entanglement entropy of Hawking radiation, which decreases and vanishes at the end of evaporation. However, as we carefully explained above, requiring the area limit means assuming new physics beyond semiclassical gravity at time scales that are much shorter than the black hole lifetime. In particular, this means that the validity of semiclassical general relativity as a low-energy effective field theory and the holographic principle are incompatible.

\subsection*{Acknowledgement}

L.~B. acknowledges financial support from the European Union’s Horizon 2020 research and innovation programme under the Marie Sklodowska-Curie Actions (grant agreement ID: 101106345-NLQG). F.~D.~F.~acknowledges financial support from the PRIMUS/-23/SCI/005 and UNCE24\-\/SCI/016 grants by Charles University, and the GAR 23-07457S grant from the Czech Science Foundation.


\bibliographystyle{utphys}
\bibliography{References}


\end{document}